\documentclass[a4paper,aps,11pt,nofootinbib,showpacs]{revtex4}

\usepackage{latexsym}
\usepackage{amsmath,amssymb}
\usepackage{graphicx,color}

\begin{document}

\title{Toroidal Spiral Nambu-Goto Strings around Higher-Dimensional Black Holes}

\today

\hfill{OCU-PHYS 321, AP-GR 72}


\author{Takahisa Igata} 
\author{Hideki Ishihara}
\affiliation{%
 Department of Mathematics and Physics,
 Graduate school of Science, Osaka City University,
 Osaka 558-8585, Japan}

\begin{abstract}
 We present solutions of the Nambu-Goto equation for test strings 
in a shape of toroidal spiral in five-dimensional spacetimes. 
In particular, we show that stationary toroidal spirals exist around 
the five-dimensional Myers-Perry black holes. 
We also show the existence of innermost stationary toroidal spirals 
around the five-dimensional black holes like geodesic particles orbiting 
around four-dimensional black holes. 
\end{abstract}

\maketitle

A lot of attention has been paid to higher-dimensional spacetimes, 
which is inspired by unified theories. 
In particular, many studies are devoted to understanding of 
physical properties of higher-dimensional black holes. 
It has been clarified that the black holes in higher dimensions have 
rich geometrical structures that have no analogue 
in four-dimensions (see \cite{LivingReview} for a review). 

The motion of a test particle is one of useful probes of geometry 
around the black holes. 
It is remarkable that a higher-dimensional generalization of 
rotating black hole, the Myers-Perry metric\cite{Myers:1986un}, 
allows separation of variables in the geodesic 
Hamilton-Jacobi equation\cite{Frolov:2002xf, Frolov:2007nt}. 
In addition, it was shown that the Myers-Perry black hole also admits 
separation of variables in the Nambu-Goto equation for stationary strings 
along the Killing time\cite{Frolov:2004qw, Kubiznak:2007ca, Ahmedov:2008pq}. 

We study, in this article, a Nambu-Goto test string 
which has a geometrical symmetry 
in the target space described by the Myers-Perry metric in five-dimensions. 
The metric admits an isometry group containing two commutable rotations. 
We assume a Killing vector which 
generates a combination of the two rotations is tangent to a worldsheet 
of a string. 
Then, on a time slice, the string has a spiral shape along a circle. 
We call it a \lq{\sl toroidal spiral string}\rq . 

Generally, if a Killing vector of a target space is tangent to a worldsheet 
of a string, the string is called a 
cohomogeneity-one string\cite{Frolov, Ishihara:2005nu}.  
Both the stationary strings, which are associated with a timelike Killing vector, 
and the toroidal spiral strings, which are defined above, 
are members of the cohomogeneity-one strings. 
The Nambu-Goto action for a cohomogeneity-one string associated with 
a Killing vector, say $\xi$, 
is reduced to the geodesic action 
\begin{align}
	S = \int \sqrt{(\xi\cdot\xi) h_{MN}dx^M dx^N}
\label{Red_Action}
\end{align}
in the orbit space of $\xi$. 
Here, $h_{MN}$ is the projection metric 
with respect to $\xi$ which is defined by 
\begin{equation}
	h_{MN} = g_{MN}- \frac{\xi_M \xi_N}{\xi\cdot\xi}, 
\label{Proj_Metric}
\end{equation}	
where $g_{MN}$ is a metric of a target space which admits the 
Killing vector $\xi$, and $\xi\cdot\xi$ is the norm of $\xi$
\cite{Frolov, Ishihara:2005nu}.
There are a lot of works on the cohomogeneity-one strings in 
a variety of contexts\cite{Burden, deVega, Frolov, Ishihara:2005nu, Ogawa}.
We consider, here, the cohomogeneity-one string as a probe of the geometry 
of higher-dimensional black holes.

A gravitational field around a compact object allows 
the existence of bounded orbits of particles in four-dimensions. 
This is easily understood by a balance of the gravitational force 
and the centrifugal force. 
In general relativity, strong gravity of a black hole forbids the existence of 
stable circular orbit inside a critical radius, 
so called, {\sl innermost stable circular orbit}. 
In contrast, there is no stable circular orbit of the geodesic particle at all 
around a higher-dimensional asymptotically flat black hole\cite{Frolov:2007nt}
\footnote{There exist stable circular orbits of particles around five-dimensional 
squashed  Kaluza-Klein black holes\cite{SQKK}.}. 
This is because the gravitational force can not be in balance with 
the centrifugal force in the higher dimensions. 

We show, in this article, that there exist stationary toroidal spiral strings 
around a higher-dimensional black hole which are achieved by the balance of 
centrifugal force and string tension in five-dimensional cases. 
Furthermore, we show appearance of innermost stationary 
toroidal spiral strings around five-dimensional black holes 
by the effect of gravity. 

Let us start with a cohomogeneity-one string in the five-dimensional Minkowski 
spacetime of the metric 
\begin{equation}
	ds^2
		=-dt^2+d\rho^2+ \rho^2d\phi^2 +d\zeta^2+ \zeta^2d\psi^2.
\label{Minkowski}
\end{equation}
We consider a toroidal spiral string, namely, 
we assume that the Killing vector of the metric 
\begin{equation}
	\xi = \partial_\phi + \alpha \partial_\psi
\label{Killing_vector}
\end{equation}
is tangent to the worldsheet of the string, 
where $\alpha$ is an arbitrary constant. 
By using  a parameter $\sigma$ on the worldsheet, 
the Killing vector is written as $\xi=\partial_\sigma$ on the worldsheet.

The dynamical system of the toroidal spiral strings is 
reduced to the system of geodesics in the metric
\begin{align}
	ds_4^2&= (\xi\cdot\xi) h_{\mu\nu}dx^\mu  dx^\nu 
\cr
		&= (\rho^2+\alpha^2\zeta^2)\left(-dt^2+ d\rho^2+d\zeta^2  \right) 
		+ \rho^2 \zeta^2 d\bar\varphi^2, 
\label{Metric_h}
\end{align}
where we have used a new variable $\bar\varphi :=\psi - \alpha \phi$. 
The action \eqref{Red_Action} in this metric is equivalent to 
\begin{align}
	S 
		=\int \left[\frac{1}{2N}
		\left((\rho^2+\alpha^2\zeta^2)\left(-\dot t^2+ \dot \rho^2
		+\dot \zeta^2  \right) 
		+ \rho^2 \zeta^2 \dot {\bar\varphi}^2\right) -\frac{N}{2}\right] d\tau, 
\label{Metric_h}
\end{align}
where the overdot denotes differentiation with respect to a parameter $\tau$, 
and $N$ is the Lagrange multiplier. 

The Hamiltonian for the particle becomes
\begin{align}
	H 
		= \frac{N}{2} \left(\frac{-p_t^2+ p_\rho^2+ p_\zeta^2 }
		{\rho^2+\alpha^2\zeta^2}
		+ \frac{p_{\bar\varphi}^2}{\rho^2 \zeta^2}  + 1\right),
\end{align}
where $p_\mu $ are the canonical momenta conjugate to $x^\mu $. 
Using the constants of motion 
$	p_t=-E$ and $p_{\bar\varphi}=L$, 
we have the reduced Hamiltonian in two-dimensions 
\begin{equation}
	H = \frac{N}{2} \left(\frac{ p_\rho^2+ p_\zeta^2 }{\rho^2+\alpha^2\zeta^2}
		+  V_{\rm eff}(\rho, \zeta) \right),
\end{equation}
where
\begin{align}
	V_\text{eff}(\rho, \zeta)
		=- \frac{E^2}{\rho^2+\alpha^2\zeta^2}+ \frac{L^2}{\rho^2 \zeta^2} +1.
\end{align}
By variation with $N$, we have the constraint equation $H\simeq 0$.

\begin{figure}
 \includegraphics[width=8cm]{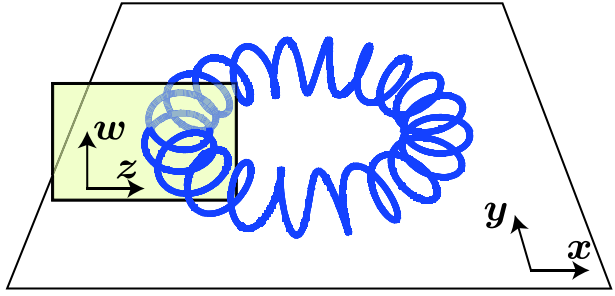}
\caption{
Snap shot of a toroidal spiral string. 
The string coils around a torus embedded in the four-dimensional 
Euclid space 
$(x, y, z, w)=(\rho\cos\phi, \rho\sin\phi, \zeta\cos\psi, \zeta\sin\psi)$.
}
\label{ToroidalSpiral}
\end{figure}

We find easily stationary solutions 
 $\rho=\rho_0=const.$ and $\zeta=\zeta_0=const.$ 
at the global minimum of $V_\text{eff}$.  
The constant radii and $E$ are given by
\begin{align}
	\rho_0 = \sqrt{\alpha L}, \quad 
	\zeta_0 = \sqrt{\frac{L}{\alpha }}, \quad \mbox{and}\quad
	E^2 = 4\alpha L.
\end{align}
In the parameter choice $\tau=t$, i.e., $N=\sqrt{\alpha L}$, and 
$\sigma=\phi$, we have
\begin{equation}
	\psi= \alpha \sigma + \sqrt{\frac{\alpha}{L}} \tau . 
\end{equation}
On a $t=const.$ surface, the string has a shape of toroidal spiral 
which lies on the two-dimensional torus
\begin{equation}
	ds_{S^1\times S^1}^2=\rho_0^2 d\phi^2+\zeta_0^2 d\psi^2 
\end{equation}
embedded in the four-dimensional Euclidean space (See Fig.\ref{ToroidalSpiral}). 
If $\alpha$ is a rational number $n_2/n_1$, where $n_1$ and $n_2$ are relatively 
prime integers, the string  coils around the torus $n_1$ times 
in $\phi$ direction while $n_2$ times in $\psi$ direction, 
then the string is closed. 
The tangent vector to the worldsheet
\begin{equation}
	\eta=\partial_\tau - \frac{1}{2\sqrt{\alpha L}}\partial_\sigma 
	=\partial_t - \frac12\left(\frac{1}{\rho_0}\partial_\phi 
		- \frac{1}{\zeta_0}\partial_\psi\right)
\end{equation} 
is a timelike Killing vector then the string is stationary. 
Since the worldsheet is spanned by two commutable 
Killing vectors $\xi$ and $\eta$, the two-dimensional surface is intrinsically 
flat. 
The angular momenta in $\phi$ and $\psi$ directions, and the total energy of 
the string are in proportion to  $-\alpha L, L$, and $E$, respectively.

Next, we consider the toroidal spiral strings in  
the five-dimensional Myers-Perry metric
\begin{align}
	ds^2 =& - dt^2
 +\frac{2 M}{\Sigma} \Big( dt-  a \sin^2\theta d\phi - b \cos^2\theta d\psi \Big)^2
\cr
	&+ \frac{\Sigma dr^2}{\Delta}  + \Sigma d\theta^2
	 + (r^2 + a^2) \sin^2\theta d\phi^2 
\cr
	&+ (r^2 + b^2) \cos^2\theta d\psi^2\ ,
\end{align}
where
\begin{align}
	&\Delta = \frac{(r^2 + a^2)(r^2 + b^2)}{r^2}- 2 M ,
\\
	&\Sigma  = r^2 + a^2 \cos^2\theta + b^2 \sin^2\theta,
\end{align}
and $a, b$, and  $M$ are parameters related to two independent rotations and mass, 
respectively. 

The metric of the orbit space for toroidal spiral strings, which is associated with 
the Killing vector \eqref{Killing_vector}, is 
\begin{align}
	ds_4^2 =& (\xi\cdot\xi) h_{\mu\nu}dx^\mu  dx^\nu  
\cr
	=& - \Big[(\xi\cdot\xi) - \frac{2 M}{\Sigma} 
	\Big((r^2 + a^2) \sin^2\theta  
		+ \alpha^2 (r^2 + b^2) \cos^2\theta\Big) \Big] dt^2
\cr
	 &+ \frac{(\xi\cdot\xi)  \Sigma}{\Delta} dr^2
	+ (\xi\cdot\xi)  \Sigma d\theta^2
	- \frac{2 M \cos^2\theta \sin^2\theta}{\Sigma} 
		\Big[ b (r^2 + a^2) - \alpha a (r^2 + b^2)\Big]
	dt d\bar \varphi
\cr
		& + \frac{\cos^2\theta \sin^2\theta}{\Sigma} 
	\Big[2 M \Big(r^2 (a^2\sin^2\theta + b^2\cos^2\theta) + a^2 b^2\Big)
 + \Sigma (r^2 + a^2)(r^2 + b^2) \Big]
	d\bar \varphi^2 , 
\label{Reduced_Metric_MP}
\end{align}
where $\bar\varphi=\psi-\alpha \phi$, 
and the norm of $\xi$ is  given by
\begin{equation}
	\xi\cdot\xi 
	= (r^2 + a^2) \sin^2\theta  + \alpha^2 (r^2 + b^2) \cos^2\theta 
		+\frac{2 M}{\Sigma} (a \sin^2\theta  + \alpha b \cos^2\theta)^2 . 
\end{equation}

\begin{figure}
 \includegraphics[width=6cm]{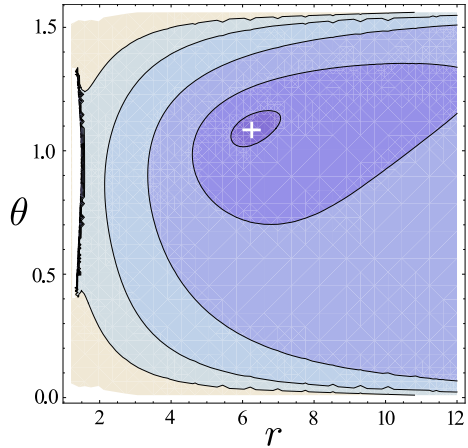}
\caption{
Contour plot of typical shape of the effective potential  
in the Myers-Perry metric.  
Parameters are chosen as  $\alpha=2, L/E=3/2$, and black hole parameters are 
$2M=1, a=1/4 , b=1/5 $, for example. 
The local minimum of the potential exists at the mark \lq +\rq\ in the figure.
}
\label{PoteffMP}
\end{figure}

Since the metric \eqref{Reduced_Metric_MP} has the Killing vectors $\partial_t$ 
and $\partial_{\bar\varphi}$, 
the corresponding canonical momenta are constants of motion. 
Setting $p_t=-E=const.$ and $p_{\bar\varphi}=L=const.$, 
we get the reduced Hamiltonian as 
\begin{align}
	H &= \frac{N}{2} \left[ (\xi\cdot\xi)^{-1} \left(
		h^{rr}p_r^2 + h^{\theta\theta}p_\theta^2\right)
		+ V_{\rm eff}(r, \theta) \right], 
\label{H_MP}
\end{align}
where
$	h^{rr} = {\Delta }/{\Sigma} $ and $
	h^{\theta\theta} = {1}/{\Sigma}$,
and
\begin{align}
	V_{\rm eff}(r, \theta)
		=& - \frac{1}{\Sigma}\left((r^2 +  2 M)+ \frac{4 M^2}{\Delta } 
		+ (a^2 \cos^2\theta + b^2 \sin^2\theta)\right) \frac{E^2}{\xi\cdot\xi}
\cr
		& + \frac{1}{\Sigma}\left(\frac{(a^2 - b^2)\Big(r^2(1 - \alpha^2)
	 + (a^2 - \alpha^2 b^2)\Big)
	 - 2 M (a - \alpha b)^2}{r^2 \Delta } 
	 +\frac{1}{\cos^2\theta}
	 +  \frac{\alpha^2}{\sin^2\theta}\right) \frac{L^2}{\xi\cdot\xi} 
\cr
	& + \frac{4 M (- b (r^2 + a^2) + \alpha a (r^2 + b^2)) }
		{r^2 \Delta \Sigma } \frac{E L}{\xi\cdot\xi}
		 + 1 . 
\label{Veff_MP}
\end{align}
The dynamical system of the toroidal spiral strings in the Myers-Perry 
black hole is reduced to the two-dimensional particle system 
specified by the Hamiltonian \eqref{H_MP} with \eqref{Veff_MP}. 
A typical potential shape is drawn as a contour plot in Fig.\ref{PoteffMP}.
There exists a stationary solution 
$r=r_0=const.$ and $\theta=\theta_0=const.$ at the local minimum 
of $V_\text{eff}(r, \theta)$. 

To inspect conditions for existence of the stationary solution, 
we restrict ourselves to the non-rotational case $a=b=0$, for simplicity. 
By use of the variables $\rho:=r\sin\theta$ and	$\zeta:=r\cos\theta$ which 
are used in Minkowski case, the effective potential \eqref{Veff_MP} becomes 
\begin{align}
	V_\text{eff}(\rho, \zeta) 
	&= \frac{E^2}{\rho^2+\alpha^2\zeta^2}
		\left( -1-\frac{2M}{\rho^2+\zeta^2- 2M}\right)
		+ \frac{L^2}{\rho^2  +\alpha^2\zeta^2}
		\left(\frac{\alpha^2}{\rho^2}+\frac{1}{\zeta^2}\right) 
		+ 1 . 
\end{align}
We can interpret the two terms in the first parenthesis 
as attractive terms by string tension and gravity, and two terms in the second 
parenthesis as centrifugal repulsion by $\phi$ and $\psi$ rotations, respectively.

\begin{figure}
\includegraphics[width=15cm]{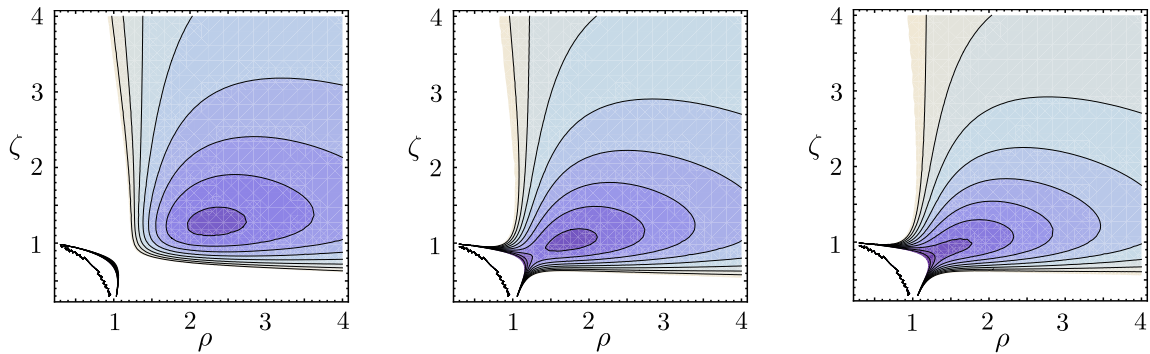}
\caption{
Contour plot of typical shapes of effective potential 
in the Schwarzschild-Tangherlini metric with $2M=1$. 
Parameters are chosen as $\alpha=2$, $L/E=0.7$(left), $L/E=0.62$(center),
and $L/E=0.6$(right). Local minima of the effective 
potential appear in the left and center panels, while the local minimum 
disappears in the right panel. 
}
\label{PoteffSch}
\end{figure}


The potential shapes are in Fig.\ref{PoteffSch}.
As same as in the Minkowski case, we can find stationary solutions
$\rho=\rho_0$ and $\zeta=\zeta_0$ at the minimum of $V_\text{eff}(\rho, \zeta)$. 
The explicit values of $\rho_0$ and $\zeta_0$ are given by solving the coupled 
algebraic equations 
\begin{equation}
 	\partial_\rho V_{\rm eff}(\rho, \zeta)=0,  \quad  
 	\partial_\zeta V_{\rm eff}(\rho, \zeta)=0,  \quad
	\mbox{and} \quad 
	V_{\rm eff}(\rho, \zeta)=0. 
\label{Minimum_Point}
\end{equation}
On a time slice $t=const.$, the stationary toroidal spiral string solution 
coils around a two-dimensional torus S$^1 \times$ S$^1$ 
of radii $\rho=\rho_0$ and $\zeta=\zeta_0$ that lies on 
S$^3$ of the radius $r_0^2=\rho_0^2+\zeta_0^2$ which is surrounding the black hole.

In contrast to the flat case, because of the gravitational term, 
the minimum of $V_\text{eff}(\rho, \zeta)$ is a local minimum 
in the black hole case. 
The local minimum in the $\rho$-$\zeta$ plane moves with 
the value of $L/E$, and disappears if 
$L/E$ becomes less than a critical value for each $\alpha$. 

\begin{figure}
 \includegraphics[width=10cm]{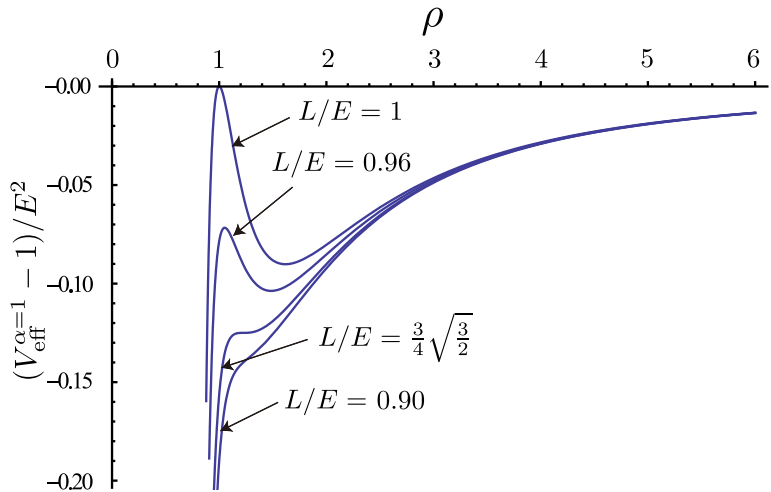}
\caption{
Cross sections of effective potential for $\alpha=1$ and various values of 
$L/E$ in the Schwarzschild-Tangherlini metric with $2M=1$.  
The local minimum disappears if $L/E=\frac{3}{4}\sqrt{\frac{3}{2}}$.
}
\label{Potential_Crosssection}
\end{figure}

\begin{figure}
 \includegraphics[width=8cm]{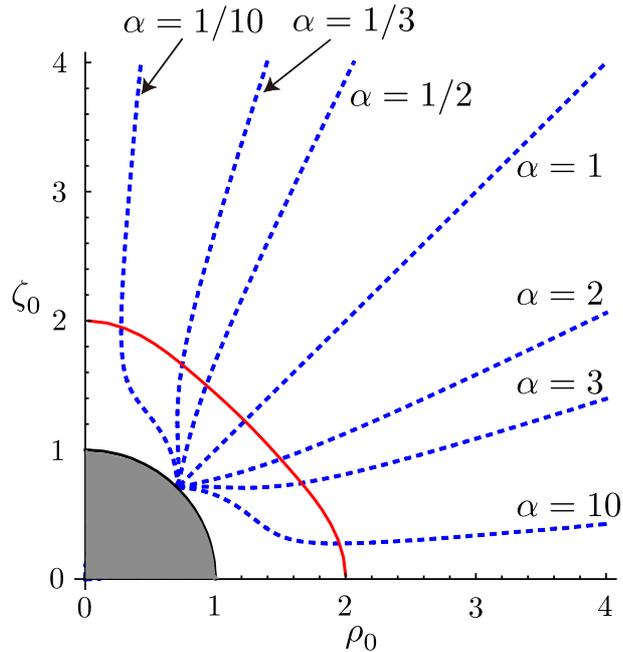}
\caption{
The radii $(\rho_0, \zeta_0)$  of innermost toroidal spiral strings 
are plotted (solid curve). 
The radii for stationary toroidal spirals stay on broken curves for 
fixed $\alpha$. Shaded region denotes inside the black hole horizon. 
}
\label{Innermost_Radii}
\end{figure}

We see from \eqref{Minimum_Point} that the local minimum points $(\rho_0, \zeta_0)$ 
for fixed $\alpha$ stay on the curve 
\begin{align}
		(\rho_0^2+\zeta_0^2)^2(\rho_0^2 -\alpha^2\zeta_0^2) 
	+4M(1-\alpha^2) { \rho_0^2\zeta_0^2} =0
\label{Minimum_Curves}
\end{align}
in the $\rho$-$\zeta$ plane. 
In the case of $\alpha=1$, $\rho_0$ is simply equal to $\zeta_0$. 
The cross section of $V_\text{eff}(\rho, \zeta)$ in $\alpha=1$ case 
along the line $\zeta=\rho$ becomes
\begin{align}
	V_\text{eff}^{\alpha=1}\Big\vert_{\zeta=\rho}(\rho) 
	&= -\frac{E^2}{2\rho^2}
		\left(1+\frac{M}{\rho^2- M}\right)
		+ \frac{L^2}{\rho^4}
		+ 1 .
\end{align}
We easily find that the innermost radii are $\rho_0=\zeta_0=\sqrt{3M}$ 
(see Fig.\ref{Potential_Crosssection}). 
For $\alpha \neq 1$ cases the curves \eqref{Minimum_Curves} and innermost 
radii of stationary toroidal spirals are shown in Fig.\ref{Innermost_Radii}.


The behavior of the toroidal spiral strings around higher-dimensional black holes 
is analogous to the geodesic particles around four-dimensional black holes. 
The toroidal spiral strings would play a role of probes to reveal geometrical 
properties of higher-dimensional black holes. 
This similarity 
suggests that 
the toroidal spirals could be captured by a black hole and 
be settled into stationary states if they lose center of mass energy 
by any processes. Because strings in higher dimensions hardly intersect 
with others\cite{Jackson:2004zg}, 
toroidal spirals could accumulate around a black hole like a cloud.

The toroidal spirals are characterized by commuting Killing vectors of 
which integral curves are closed in the target space. 
The stationary toroidal spirals are realized essentially by the balance 
of the string tension and the centrifugal force, it is straight forward to 
extend the solution in the case of black holes with more than five-dimensions. 
It is also possible to consider more complicated spiralling strings 
generated by more numbers of Killing vectors. 
Moreover, we can consider the black rings 
as play grounds of the toroidal spiral strings. 

It is interesting problem to study the gravitational field around 
the toroidal spirals. 
Investigation of gravitational wave emission from the toroidal spirals 
would be important. 
In four-dimensions, it is well known that cusps commonly appear 
in the evolution of closed strings\cite{Kibble:1982cb}. 
However, even if strings are closed in five-dimensions, 
as has been shown in this paper, 
they can be the stationary toroidal spirals without cusp. 
The difference comes from the dimensionality 
of the target space of the strings. 
It would be natural that the stationary toroidal spirals emit gravitational waves 
almost constantly in a long duration with periodic waveforms\cite{Ogawa:2008yx}   
in the higher-dimensional universe. 

The toroidal spirals are extended sources of gravity which have two 
independent rotations. They would mimic gravitational fields of black rings 
in far regions. 
Toroidal spirals in six or more dimensions would supply hints to 
construct black ring solutions\cite{Emparan:2007wm}.

Though we mainly focus on the stationary solution of toroidal spirals 
in this article, investigation of the dynamics of toroidal spirals is 
a tractable issue. 
We will report integrability of the system in a separate paper\cite{Igata-Ishihara}. 

This work is supported by the Grant-in-Aid for Scientific Research No.19540305.



\begin{thebibliography}{99}

\bibitem{LivingReview}
R. Emparan and H. S. Reall,
"Black Holes in Higher Dimensions",
Living Rev. Relativity 11,  (2008),  6. URL (cited on 30 Oct. 2009):
http://www.livingreviews.org/lrr-2008-6 . 

\bibitem{Myers:1986un}
  R.~C.~Myers and M.~J.~Perry,
  Annals Phys.\  {\bf 172}, 304 (1986).

\bibitem{Frolov:2002xf}
  V.~P.~Frolov and D.~Stojkovic,
  Phys.\ Rev.\  D {\bf 67}, 084004 (2003);
Ibid. D {\bf 68}, 064011 (2003) .

\bibitem{Frolov:2007nt}
  V.~P.~Frolov and D.~Kubiznak,
  Phys.\ Rev.\ Lett.\  {\bf 98}, 011101 (2007) .

%
\bibitem{Frolov:2004qw}
  V.~P.~Frolov and K.~A.~Stevens,
  Phys.\ Rev.\  D {\bf 70}, 044035 (2004) .


\bibitem{Kubiznak:2007ca}
  D.~Kubiznak and V.~P.~Frolov,
  JHEP {\bf 0802}, 007 (2008) .

\bibitem{Ahmedov:2008pq}
  H.~Ahmedov and A.~N.~Aliev,
  Phys.\ Rev.\  D {\bf 78}, 064023 (2008) .


\bibitem{Frolov}
         V.~P.~Frolov, V.~Skarzhinsky, A.~Zelnikov and O.~Heinrich,
         Phys.\ Lett.\  B {\bf 224}, 255 (1989).
\\
         V.~P.~Frolov, S.~Hendy and J.~P.~De Villiers,
         Class.\ Quant.\ Grav.\  {\bf 14}, 1099 (1997).


 \bibitem{Ishihara:2005nu}
         H.~Ishihara and H.~Kozaki,
         Phys.\ Rev.\  D {\bf 72}, 061701 (2005).

\bibitem{Burden}
         C.~J.~Burden and L.~J.~Tassie,
         Austral.\ J.\ Phys.\  {\bf 35}, 223 (1982); 
Ibid. {\bf 37}, 1 (1984); \\
  C.~J.~Burden,
  Phys.\ Rev.\  D {\bf 78}, 128301 (2008).

\bibitem{deVega}
         H.~J.~de Vega, A.~L.~Larsen and N.~G.~Sanchez,
         Nucl.\ Phys.\  B {\bf 427}, 643 (1994); \\
         A.~L.~Larsen and N.~G.~Sanchez,
         Phys.\ Rev.\  D {\bf 50}, 7493 (1994); 
%
Ibid. D {\bf 51}, 6929 (1995); \\
%
         H.~J.~de Vega and I.~L.~Egusquiza,
         Phys.\ Rev.\  D {\bf 54}, 7513 (1996). 
%
\bibitem{Ogawa}
         K.~Ogawa, H.~Ishihara, H.~Kozaki, H.~Nakano and S.~Saito,
         Phys.\ Rev.\ D {\bf 78}, 023525 (2008); \\
%
         T.~Koike, H.~Kozaki and H.~Ishihara,
         Phys.\ Rev.\  D {\bf 77}, 125003 (2008); \\
%
  H.~Kozaki, T.~Koike and H.~Ishihara,
  arXiv:0907.2273 [gr-qc].


\bibitem{SQKK}
  H. Ishihara and K. Matsuno, Prog. Theor. Phys. {\bf 116}, 417 (2006); \\
  K.~Matsuno and H.~Ishihara,
  arXiv:0909.0134 [hep-th].


\bibitem{Jackson:2004zg}
         M.~G.~Jackson, N.~T.~Jones and J.~Polchinski,
         JHEP {\bf 0510}, 013 (2005). 



\bibitem{Kibble:1982cb}
  T.~W.~B.~Kibble and N.~Turok,
  Phys.\ Lett.\  B {\bf 116}, 141 (1982).



 \bibitem{Ogawa:2008yx}
         K.~Ogawa, H.~Ishihara, H.~Kozaki and H.~Nakano,
         Phys.\ Rev.\  D {\bf 79}, 063501 (2009). 

\bibitem{Emparan:2007wm}
  R.~Emparan, T.~Harmark, V.~Niarchos, N.~A.~Obers and M.~J.~Rodriguez,
  JHEP {\bf 0710}, 110 (2007). 


\bibitem{Igata-Ishihara}
 T.Igata and H.Ishihara, in preparation.
 
\end{thebibliography}
\end{document}